\documentclass[letterpaper,british,english,reprint, aps]{svjour3}
\usepackage[T1]{fontenc}
\usepackage{color}
\usepackage{array}
\usepackage{mathrsfs}
\usepackage{url}
\usepackage{amsmath}
\usepackage{amssymb}
\usepackage{stmaryrd}
\usepackage{graphicx}
\usepackage[numbers]{natbib}

\makeatletter

\pdfpageheight\paperheight
\pdfpagewidth\paperwidth

\newcommand*\LyXZeroWidthSpace{\hspace{0pt}}
\providecommand{\tabularnewline}{\\}

\usepackage{geometry}
\usepackage{tikz}
\usetikzlibrary{quantikz}
\newcommand{\ls}{\lstick}
\newcommand{\rs}{\rstick}
\newcommand{\gt}{\gate}
\newcommand{\gh}[1]{\gate[#1,disable auto height]}
\newcommand{\qb}{\qwbundle}
\newcommand{\se}{\slice}

\@ifundefined{showcaptionsetup}{}{%
 \PassOptionsToPackage{caption=false}{subfig}}
\usepackage{subfig}
\makeatother

\usepackage{babel}
\begin{document}
\title{\textcolor{black}{Unitary (semi)causal quantum-circuit representation
of black hole evaporation}}
\author{Bogus\l aw Broda}
\institute{Department of Theoretical Physics, Faculty of Physics and Applied
Informatics, University of \L \'{o}d\'{z}, 90-236 \L \'{o}d\'{z},
Pomorska 149/153, Poland\\
ORCID: 0000-0001-9753-7207\\
\texttt{boguslaw.broda@uni.lodz.pl}\foreignlanguage{british}{\texttt{}}~\\
\foreignlanguage{british}{\texttt{https}}\texttt{://}\foreignlanguage{british}{\texttt{www}}\texttt{.uni.lodz.pl/en/employee/boguslaw-broda}}
\date{21 February 2024}
\maketitle
\begin{abstract}
A general structure of unitary evolution (evaporation) of the black
hole, respecting causality imposed by the event horizon (semicausality),
has been derived and presented in the language of quantum circuits.
The resulting consequences for the evolution of the corresponding
entanglement entropy and the entropy curve have been determined. As
an illustration of the general scheme, two families of qubit toy models
have been discussed: tensor product models and controlled non-product
models.
\end{abstract}

\section{Introduction}

The black hole (BH) information paradox formulated in \citep{Hawking1976}
is still an interesting, unsolved, and inspiring problem. Recent progress,
starting with the two famous papers made in 2019, \citep{Almheiri2019}
and \citep{Penington2020}, consists in the computation of the von
Neumann entropy of the radiation emitted by BHs in the framework of
some specific models of quantum gravity. It appears that in these
models, the shape of the entropy curve is essentially given by the
Page curve \citep{Page1993}. Instead, in the present work, we concentrate
on the description of the unitary evolution and evaporation process
of BHs directly in terms of qubits, according to the ideas reviewed
by e.g.\ \citep{Avery2013}. Extensive literature with some 50 references
to various qubit models is given in \citep{Osuga2018} (see also proposals
in \citep{Broda2020,Broda2021,Broda2021PLB}). Despite these large
studies, it seems that the important issue of properly understood
causality in the process of transport of quantum information in the
presence of the BH event horizon has been largely ignored. The aim
of our work is to at least partially fill the gap.

The paper consists of two parts, and the whole analysis is performed
in the language of quantum circuits. In the first part (Section \ref{sec:General-structure-of}),
we implement the concept of causality that is, in our opinion, most
natural for the description of the process of evolution (and evaporation)
of a BH. Namely, taking into account the role played by the BH event
horizon, the kind of causality appropriate in this case is given by
the (already existent in literature) notion of \emph{semicausality}.
The semicausality in turn imposes some restrictions on the structure
of unitary evolution and, in consequence, on the evolution of entropy
and (in a graphic presentation) on the entropy curve. In the second
part of the paper (Section \ref{sec:Some-examples-of}), we present
and analyze two large families of unitary qubit toy models that explicitly
implement the idea of (semi)causality. The first family comprises
some tensor product models, whereas the second one is a non-product
generalization of the first family with a small number of control
qubits.

\section{The general structure of unitary semicausal circuits\label{sec:General-structure-of}}

In this section, we would like to establish a general structure of
unitary transformations governing the evolution of evaporating BHs,
which properly respect causality implied by the presence of the BH
event horizon. Because of our later needs (see Section \ref{sec:Some-examples-of}),
the answer will be formulated in the language of quantum circuits.
Since unitary transformations (and their corresponding quantum circuits)
are supposed to act in regions both outside and inside the BH, restrictions
following from causality should be properly encoded in the structure
of the circuit. It appears that the version of causality appropriate
in the context of a BH is semicausality. In quantum mechanics, the
notion of semicausality has a short history. The concept of semicausality
first appeared in \citep{Beckman2001}, was analyzed in \citep{Eggeling2002}
(see also \citep{Piani2006}), and was later discussed by \citep{Schumacher2012}
(see also \citep{Schumacher2005}) in the circuit (diagrammatic) representation.
One should also note that semicausality has already explicitly appeared
in the context of BHs in \citep{Braunstein2018}.

Occasionally, we loosely use the term \emph{(semi)causality} instead
of semicausality (e.g., in the title) or even causality, but referring
to causality in the present paper, we exclusively mean the notion
of semicausality (precisely, causality is something different\textemdash it
is semicausality ``valid in both directions'').

\subsection{Semicausality in quantum circuits}

\begin{figure}[h]
\centering{}\hspace*{\fill}\subfloat[]{\begin{centering}
\begin{quantikz}
\ls{B} & \qw                      & \gh{2}{U^{(\textrm BX)}} & \qw \\
\ls{X} & \gh{2}{U^{(\textrm XA)}} & \qw                      & \qw \\
\ls{A} &                          & \qw                      & \qw
\end{quantikz}
\par\end{centering}
\label{fig:Schumacher}}\hspace*{\fill}\subfloat[]{\begin{centering}
\begin{quantikz}
\ls{B} & \gt{U^{(\textrm B)}}          & \gh{2}{U^{(\textrm BX)}}   & \qw \\
\ls{X} & \gh{2}{U^{(\textrm XA)}}  	& \qw      				  & \qw \\
\ls{A} &           					& \gt{U^{(\textrm A)}}       & \qw
\end{quantikz}
\par\end{centering}
\label{fig:extSchumacher}}\hspace*{\fill}\caption{Diagrammatic representation of the operator $U^{\left(\textrm{BXA}\right)}$
defining the unitary evolution of the tripartite system $\textrm{BXA}$,
where no signalling from Bob (B) to Alice (A) is allowed (semicausality).
Transmission of information can only be related to particle(s) (X)
sent from Alice (A) to Bob (B), but not in the other direction. Here
(a) presents the original version of the operator, given in (\ref{eq:U(BXA)_tensored}),
whereas (b) presents its slightly generalized version (\ref{eq:U(BXA)_tensored_an}),
which is better suited for our further considerations.}
\label{fig:Schumacher-extSchumacher}
\end{figure}
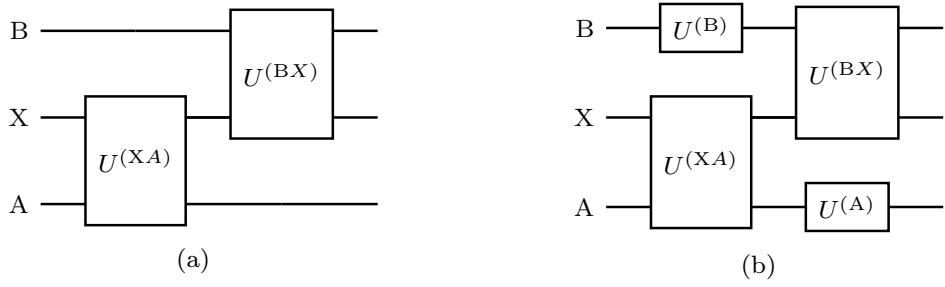

An inherent characteristic property of the BH is its one-way transmission
of information (and particles) through its event horizon. This property
perfectly harmonizes with the notion of semicausality introduced in
the context of quantum mechanics and is phrased in the following way
\citep{Beckman2001}: If a bipartite operation $\mathcal{E}$ does
not enable superluminal signalling from Bob to Alice (denoted in literature
as $\textrm{B}\nrightarrow\textrm{A}$, or similarly), then we say
that $\mathcal{E}$ is semicausal. In the interesting context of the
event horizon, Bob should be placed inside the BH, whereas Alice should
reside outside the BH. The main result of \citep{Eggeling2002} is
a proof of a conjecture by DiVincenco (mentioned by \citep{Beckman2001}),
which (in physics terms) says that if a device allows no signalling
from Bob to Alice (semicausality), we can represent it explicitly
as a device involving possibly a particle sent from Alice to Bob but
none in the other direction. More precisely (in mathematical terms),
let the unitary evolution of the tripartite system $\textrm{BXA}$
be described by the operator $U^{\left(\textrm{BXA}\right)}$. Then
$U^{\left(\textrm{BXA}\right)}$ can be represented by the following
product containing unitary operators $U^{\left(\textrm{BX}\right)}$
and $U^{\left(\textrm{XA}\right)}$ \citep{Schumacher2005}:
\begin{equation}
U^{\left(\textrm{BXA}\right)}=\left(U^{\left(\textrm{BX}\right)}\varotimes1^{\left(\textrm{A}\right)}\right)\left(1^{\left(\textrm{B}\right)}\otimes U^{\left(\textrm{XA}\right)}\right),\label{eq:U(BXA)_tensored}
\end{equation}
where $\textrm{B}$ denotes Bob's subsystem, $\textrm{X}$ is a ``particles'
sent'' subsystem, and $\textrm{A}$ denotes Alice's subsystem. A
diagrammatic (circuit in block form) presentation of (\ref{eq:U(BXA)_tensored})
is depicted in Figure \ref{fig:Schumacher} (see \citep{Schumacher2012}).
We can also present another, essentially equivalent version of the
theorem, which is better suited for our further considerations:
\begin{equation}
U^{\left(\textrm{BXA}\right)}=\left(U^{\left(\textrm{BX}\right)}\varotimes U^{\left(\textrm{A}\right)}\right)\left(U^{\left(\textrm{B}\right)}\otimes U^{\left(\textrm{XA}\right)}\right),\label{eq:U(BXA)_tensored_an}
\end{equation}
where the additional unitary operators $U^{\left(\textrm{A}\right)}$
and $U^{\left(\textrm{B}\right)}$ only redefine the subsystems $\textrm{A}$
and $\textrm{B}$ after and before the evolution (\ref{eq:U(BXA)_tensored}),
respectively. A diagrammatic version of (\ref{eq:U(BXA)_tensored_an})
is presented in Figure \ref{fig:extSchumacher}.

\begin{figure}[h]
\begin{centering}
\begin{quantikz}
\ls{$X_1$}	&\gt{U^{\left(X_1\right)}}	  &\gh{2}{U^{\left(X_1X_{2}\right)}} &\ \ldots\ \qw	  &\gh{4}{U^{\left(X_1X_2\ldots X_{N-1}\right)}} &\qw	\\
\ls{$X_2$}	&\gh{4}{U^{\left(X_2\ldots X_N\right)}}   &		 &\ \ldots\ \qw	  &		 &\qw	\\
\ls{$X_3$}    &		   &\gh{3}{U^{\left(X_3\ldots X_N\right)}} &\ \ldots\ \qw	  &		 &\qw	\\
\ls{\raisebox{1.5ex}{$\vdots$\hskip.5em}} &\qb{}  	&\qb{}	&\ \ldots\ \qb{}	 &\qb{}	&\qb{}  \\
\ls{$X_N$}	&		   &		 &\ \ldots\ \qw	  &\gt{U^{\left(X_N\right)}}    &\qw	  
\end{quantikz}
\par\end{centering}
\centering{}\caption{Diagrammatic representation of the operator $U^{\left(X_{1}X_{2}\ldots X_{N}\right)}$,
which is an arbitrarily large iteration of the operator $U^{\left(\textrm{BXA}\right)}$
defined in (\ref{eq:U(BXA)_tensored_an}) and depicted in Figure \ref{fig:extSchumacher}.}
\label{fig:iterSchumacher}
\end{figure}
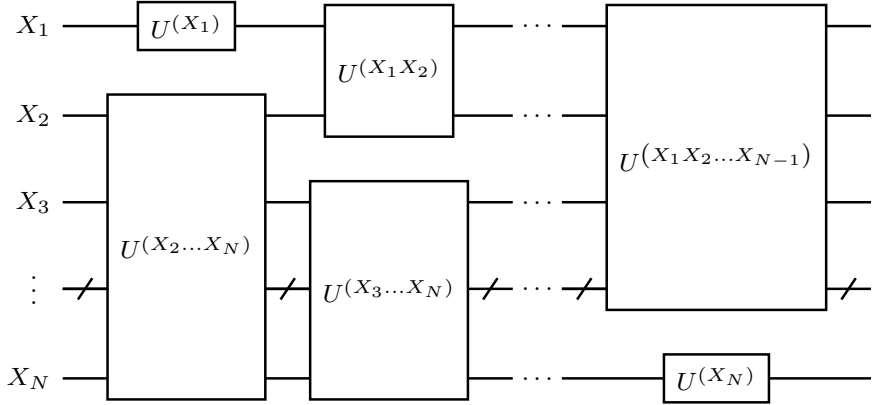

In the process of BH evaporation (and BH evolution in general), we
deal with a continuous flux of particles (e.g., external real ``matter''
particles and/or Hawking's virtual ones) constantly crossing the event
horizon and entering the BH. Therefore, the formula (\ref{eq:U(BXA)_tensored})
or (\ref{eq:U(BXA)_tensored_an}) should be appropriately generalized,
i.e., iterated. Namely, the unitary evolution of the $N$-partite
system $X_{1}X_{2}\ldots X_{N}$ described by the operator $U^{\left(X_{1}X_{2}\ldots X_{N}\right)}$
should be a (huge, in the BH case) iteration of (\ref{eq:U(BXA)_tensored_an}),
assuming the form 

\begin{equation}
\begin{aligned}U^{\left(X_{1}X_{2}\ldots X_{N}\right)}= & \left(U^{\left(X_{1}X_{2}\ldots X_{N-1}\right)}\otimes U^{\left(X_{N}\right)}\right)\\
 & \ldots\left(U^{\left(X_{1}X_{2}\right)}\otimes U^{\left(X_{3}\ldots X_{N}\right)}\right)\left(U^{\left(X_{1}\right)}\otimes U^{\left(X_{2}\ldots X_{N}\right)}\right)
\end{aligned}
\label{eq:U(many_X)_tensored}
\end{equation}
(see Figure \ref{fig:iterSchumacher}), where all the operators $U^{(\cdots)}$
are unitary.

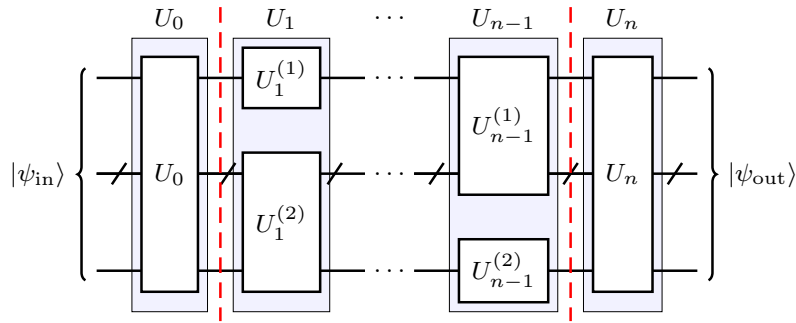
\begin{figure}[h]
\begin{centering}
\def\gg#1#2{\gategroup[{#1},steps={#2},
style={very thin,fill=blue!5,inner sep=0em},background]}
\def\sp{0em}
\begin{quantikz}
&&&\dots&&&\\[-1.5ex]
\lstick[wires=3]{$\ket{\psi_{\textrm{in}}}$}&\gh{3}{U_0}\gg{3}{1}{$U_0$}
\se{}
&[\sp]\gt{U_1^{(1)}}
\gg{3}{1}{$U_1$}
&\ \ldots\ \qw &\gh{2}{U_{n-1}^{(1)}}\gg{3}{1}{$U_{n-1}$}
\se{}
&[\sp]\gh{3}{U_n}\gg{3}{1}{$U_n$} &\qw\rstick[wires=3]{$\ket{ \psi_{\textrm{out}}}$} \\
&\qb{}   &\gh{2}{U_1^{(2)}}\qb{}&\ \ldots\ \qb{}&\qb{}      &\qb{}    &\qb{}\\
&        &	        &\ \ldots\ \qw  &\gt{U_{n-1}^{(2)}}     &         &\qw
\end{quantikz}
\par\end{centering}
\centering{}\caption{(Color online) Diagrammatic representation of the unitary evolution
operator $U$ defined in (\ref{eq:U(many_x)_tensored_generalized}),
which is a slight generalization of the operator $U^{\left(x_{1}x_{2}\ldots x_{n}\right)}$
in (\ref{eq:U(many_x)_tensored}), interpreted as the most fine-grained
particle version of the operator $U^{\left(X_{1}X_{2}\ldots X_{N}\right)}$
given in (\ref{eq:U(many_X)_tensored}) and depicted in Figure \ref{fig:iterSchumacher}.
The vertical dashed left line and the right line denote the theoretically
earliest moment and the latest one, when the BH is allowed to exist,
respectively.}
\label{fig:iterSchumacherSym}
\end{figure}

Going from $N$-partite (hugely multi-particle, in general) quantum
subsystems $X_{1},X_{2},\:$\LyXZeroWidthSpace$\ldots,X_{N}$ of
the system $X$ (here the symbol $X$ should not be confused with
a ``particles' sent'' subsystem X from the beginning of this subsection)
to $n$ qubits $x_{1},x_{2},\ldots,x_{n}$ (usually $n\gg N$), we
can replace (\ref{eq:U(many_X)_tensored}) with its (most) fine-grained
version
\begin{equation}
\begin{aligned}U^{\left(x_{1}x_{2}\ldots x_{n}\right)}= & \left(U^{\left(x_{1}x_{2}\ldots x_{n-1}\right)}\otimes U^{\left(x_{n}\right)}\right)\\
 & \ldots\left(U^{\left(x_{1}x_{2}\right)}\otimes U^{\left(x_{3}\ldots x_{n}\right)}\right)\left(U^{\left(x_{1}\right)}\otimes U^{\left(x_{2}\ldots x_{n}\right)}\right),
\end{aligned}
\label{eq:U(many_x)_tensored}
\end{equation}
or equivalently (in a shorter notation) by 
\begin{equation}
U=\left(U_{n-1}^{\left(1\right)}\otimes U_{n-1}^{\left(2\right)}\right)\ldots\left(U_{2}^{\left(1\right)}\otimes U_{2}^{\left(2\right)}\right)\left(U_{1}^{\left(1\right)}\otimes U_{1}^{\left(2\right)}\right)\equiv U_{n-1}\ldots U_{2}U_{1},\label{eq:U(many_x)_tensored_shorter}
\end{equation}
where $U_{k}\equiv U_{k}^{\left(1\right)}\otimes U_{k}^{\left(2\right)}$
($k=1,\ldots n-1$). The expansion (\ref{eq:U(many_x)_tensored})
or equivalently (\ref{eq:U(many_x)_tensored_shorter}) can be a bit
further generalized to

\begin{equation}
\begin{aligned}U= & U_{n}\left(U_{n-1}^{\left(1\right)}\otimes U_{n-1}^{\left(2\right)}\right)\ldots\left(U_{2}^{\left(1\right)}\otimes U_{2}^{\left(2\right)}\right)\left(U_{1}^{\left(1\right)}\otimes U_{1}^{\left(2\right)}\right)U_{0}\\
\equiv & U_{n}U_{n-1}\ldots U_{2}U_{1}U_{0},
\end{aligned}
\label{eq:U(many_x)_tensored_generalized}
\end{equation}
where the ``external'' unitary operators $U_{0}$ and $U_{n}$ have
been added in analogy to the generalization (\ref{eq:U(BXA)_tensored_an}),
and they only redefine the \emph{in }(initial) state $\left|\psi_{\textrm{in}}\right\rangle $
and the \emph{out} (final) state $\left|\psi_{\textrm{out}}\right\rangle $,
respectively (see Figure \ref{fig:iterSchumacherSym}).
In the context of BH evolution, the operators $U_{0}$ and $U_{n}$
are supposed to act before the BH forms and after its complete evaporation,
respectively.

One should stress that according to (\ref{eq:U(many_x)_tensored_shorter}),
the decomposition $\mathcal{H}=\mathcal{H}_{k}^{\left(1\right)}\otimes\mathcal{H}_{k}^{\left(2\right)}$
($k=1,\ldots,n-1$) of the total Hilbert space $\mathcal{H}$ onto
a product of the two Hilbert spaces $\mathcal{H}_{k}^{\left(a\right)}$,
$a=1,2$, where the superscripts 1 and 2 correspond to the interior
and the exterior of the BH, respectively, is ``dynamical''. In other
words, the decomposition of $\mathcal{H}$ gradually changes in the
course of the (indexed by $k$) evolution of the BH, as depicted in
Figures \ref{fig:iterSchumacherSym} and \ref{fig:between2U}. This
observation undermines the validity of the fixed global decomposition
of the total Hilbert space $\mathcal{H}$ in the context of BH evaporation.

\subsection{Bounds on entanglement entropy\label{subsec:Bounds-on-entanglement}}

\begin{figure}[h]
\begin{centering}
\begin{center}
\begin{quantikz}
& U_{k-1} &&&&& U_k &\\[-2ex]
\ls{$\ket{x_1}$} &\gh{2}{U_{k-1}^{(1)}\in\mathcal{H}_{k-1}^{\left(1\right)}\qquad}
\gateoutput[wires=2]{B}
\se{$\mathcal{S}_{k}^{-}\dbinom{\rho_{k}^{(1)-}}{\rho_{k}^{(2)-}}$}
&\qw&\phantomgate{X}\qw\se{$\ket{\psi_k}$} &\qw &\phantomgate{X}\qw\se{$\mathcal{S}_{k}^{+}\dbinom{\rho_{k}^{(1)+}}{\rho_{k}^{(2)+}}$} &\gh{3}{\qquad U_k^{(1)}\in\mathcal{H}_k^{\left(1\right)}} \gateinput[wires=2]{B}   &\qw &\qw\\
\ls{\raisebox{1.5ex}{$\vdots$\hskip.5em}}&\qb{} &\qw&\qb{} &\qw &\qb{} 	&\qw 	 &\qw&\qb{} \\
\ls{$\ket{x_k}$} &\gh{3}{U_{k-1}^{(2)}\in\mathcal{H}_{k-1}^{\left(2\right)}\qquad}\gateoutput[wires=1]{X}  &\qw&\qw &\qw &\qw &\gateinput[wires=1]{X}  &\qw&\qw \\
\ls{\raisebox{1.5ex}{$\vdots$\hskip.5em}}   &\qb{}\qquad	\gateoutput[wires=2]{A}	  &\qw&\qb{}	 &\qw &\qb{}     &\gh{2}{\qquad U_k^{(2)}\in\mathcal{H}_k^{\left(2\right)}}\qw\gateinput[wires=2]{A}&\qw &\qb{} \\
\ls{$\ket{x_n}$} &			   &\qw&\qw	   &\qw &\qw	  &	&\qw&\qw
\end{quantikz}
\par\end{center}
\par\end{centering}
\centering{}\caption{(Color online) Description of the configuration between two arbitrary
\foreignlanguage{british}{neighbouring} unitary blocks of Figure \ref{fig:iterSchumacherSym}.
The (pure) state between the two blocks $U_{k-1}$ and $U_{k}$ is
denoted by $\ket{\psi_{k}}$. In our notation, the correspondence
between the reduced density matrices and the subsystems is as follows:
$\rho_{k}^{(1)-}$\textemdash $\,\left(\textrm{B}\right)\equiv\left(x_{1},\ldots,x_{k-1}\right)$,
$\rho_{k}^{(2)-}$\textemdash $\,\left(\textrm{XA}\right)\equiv\left(x_{k},\ldots,x_{n}\right)$,
$\rho_{k}^{(1)+}$\textemdash $\,\left(\textrm{BX}\right)\equiv\left(x_{1},\ldots,x_{k}\right)$,
and $\rho_{k}^{(2)+}$\textemdash $\,\left(\textrm{A}\right)\equiv\left(x_{k+1},\ldots,x_{n}\right)$
(see (\ref{eq:U(BXA)_tensored_an}) and (\ref{eq:U(many_x)_tensored})).
The entanglement entropy $\mathcal{S}_{k}^{\mp}$ is by definition
the von Neumann entropy for the reduced density matrix $\rho_{k}^{\left(a\right)\mp}$
($a=1,2$), where, as it turns out, $a$ can be arbitrary and thus
finally skipped.}
\label{fig:between2U}
\end{figure}
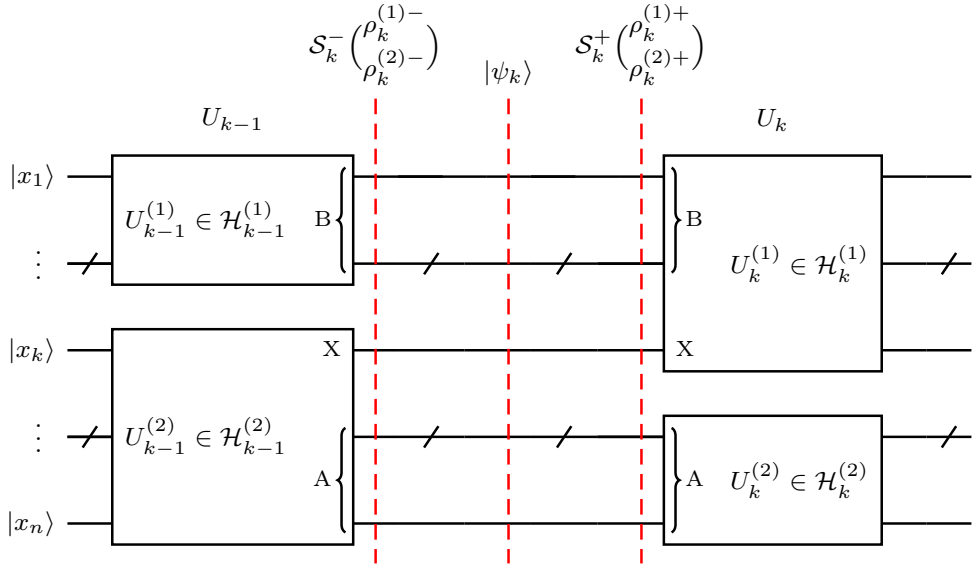

The general situation and the area of our interest between two arbitrary
neighbouring unitary blocks $U_{k-1}$ and $U_{k}$ have been presented
in Figure \ref{fig:between2U}.
The total (pure) state exiting the unitary block $U_{k-1}$ and next
entering $U_{k}$ has been denoted by $\left|\psi_{k}\right\rangle $.
As follows from Figure \ref{fig:between2U}, the configuration in
the area between the two pairs of unitary operators, $U_{k-1}\equiv U_{k-1}^{\left(1\right)}\otimes U_{k-1}^{\left(2\right)}$
and $U_{k}\equiv U_{k}^{\left(1\right)}\otimes U_{k}^{\left(2\right)}$,
gives rise to the four density operators $\rho_{k}^{(a)\mp}$, $a=1,2$.
Here the signs ``$-$'' and ``$+$'' denote the LHS and the RHS
of the area in Figure \ref{fig:between2U}, respectively, and the
values of the superscript $a$ correspond to the values $1$ and $2$
in the two decompositions of the Hilbert space $\mathcal{H}=\mathcal{H}_{k-1}^{\left(1\right)}\otimes\mathcal{H}_{k-1}^{\left(2\right)}$
and $\mathcal{H}=\mathcal{H}_{k}^{\left(1\right)}\otimes\mathcal{H}_{k}^{\left(2\right)}$
for the LHS and the RHS, respectively. The entanglement entropy $\mathcal{S}$
is the von Neumann entropy of the reduced density matrix calculated
for one of the four subsystems, which in the block notation of Figure
\ref{fig:extSchumacher} could be denoted by the superscripts $\textrm{B}$,
$\textrm{XA}$, $\textrm{BX}$, and $\textrm{A}$, and which in the
current notation of Figure \ref{fig:between2U} are denoted by the
superscripts $\left(1\right)-$, $\left(2\right)-$, $\left(1\right)+$,
and $\left(2\right)+$, respectively. Utilizing a well-known equality
between the von Neumann entropies in a two-partite division (here
denoted by the superscript $a=1,2$) of a system in a pure state $\left|\psi_{k}\right\rangle $
we can skip (on both sides, i.e., for both superscripts ``$-$''
and ``$+$'') the superscript $a=1,2$, i.e.,

\begin{equation}
\begin{aligned}\mathcal{S}\left(\rho_{k}^{\left(1\right)\mp}\right)= & -\textrm{Tr \ensuremath{\left[\rho_{k}^{\left(1\right)\mp}\ln\rho_{k}^{\left(1\right)\mp}\right]}}\\
= & -\textrm{Tr \ensuremath{\left[\rho_{k}^{\left(2\right)\mp}\ln\rho_{k}^{\left(2\right)\mp}\right]}=\ensuremath{\mathcal{S}\left(\rho_{k}^{\left(2\right)\mp}\right)}\ensuremath{\mathcal{\equiv S}\left(\rho_{k}^{\mp}\right)}}\equiv\mathcal{S}_{k}^{\mp}.
\end{aligned}
\label{eq:entropy}
\end{equation}

By assumption, the unitary evolution (\ref{eq:U(many_x)_tensored_generalized})
is entirely general and only restricted by a single kinematic condition:
semicausality. It appears that even though we should not expect too
much without any additional dynamic input, some very general results
can be derived. Namely, the global (semi)causal structure of the total
unitary evolution $U$ defined in (\ref{eq:U(many_x)_tensored_generalized})
(see Figure \ref{fig:iterSchumacherSym}) implies bounds on the evolution
of the entanglement entropy $\mathcal{S}$. The evolution of $\mathcal{S}$
itself can be characterized by the steps (here denoted by $\mathcal{S}_{k}^{\mp}$)

\begin{equation}
\mathcal{S}_{1}^{-}\left(=0\right)\rightarrow\mathcal{S}_{1}^{+}\rightarrow\mathcal{S}_{2}^{-}\rightarrow\mathcal{S}_{2}^{+}\rightarrow\cdots\rightarrow\mathcal{S}_{k}^{-}\rightarrow\mathcal{S}_{k}^{+}\rightarrow\cdots\rightarrow\mathcal{S}_{n-1}^{+}\rightarrow\mathcal{S}_{n}^{-}\rightarrow\mathcal{S}_{n}^{+}\left(=0\right),\label{eq:EntropySteps}
\end{equation}
where the presence of zeros ($\mathcal{S}_{1}^{-}=\mathcal{S}_{n}^{+}=0$)
should be obvious by virtue of Figure \ref{fig:iterSchumacherSym}.
In turn, the bounds on entropy can be derived from the Araki\textendash Lieb
version of the triangle inequality \citep{Araki1970}
\begin{equation}
\left|\mathcal{S}^{\left(\textrm{B}\right)}-\mathcal{S}^{\left(\textrm{X}\right)}\right|\leq\mathcal{S}^{\left(\textrm{BX}\right)}\leq\mathcal{S}^{\left(\textrm{B}\right)}+\mathcal{S}^{\left(\textrm{X}\right)},\label{eq:Trianle_ineq}
\end{equation}
here written in the notation of Figures \ref{fig:Schumacher} or \ref{fig:extSchumacher}.
In particular, for a one-qubit $\textrm{X}$, the evolution of the
von Neumann entropy is constrained by the inequality (essentially,
a version of the triangle inequality)
\begin{equation}
\left|\Delta\mathcal{S}\right|\equiv\left|\mathcal{S}^{\left(\textrm{BX}\right)}-\mathcal{S}^{\left(\textrm{B}\right)}\right|\leq\ln2.\label{eq:DeltaSIneq}
\end{equation}
Explicitly, (\ref{eq:DeltaSIneq}) is a consequence of the following
three inequalities:
\begin{equation}
\begin{cases}
1^{\textrm{st}} & \mathcal{S}^{\left(\textrm{BX}\right)}-\mathcal{S}^{\left(\textrm{B}\right)}\leq\mathcal{S}^{\left(\textrm{X}\right)};\\
2^{\textrm{nd}} & -\mathcal{S}^{\left(\textrm{X}\right)}\leq\mathcal{S}^{\left(\textrm{BX}\right)}-\mathcal{S}^{\left(\textrm{B}\right)}\\
3^{\textrm{rd}} & \mathcal{S}^{\left(\textrm{X}\right)}\leq\ln2,
\end{cases};\label{eq:eqDeltaSIneq}
\end{equation}
where the $1^{\textrm{st}}$ inequality follows from the RHS of the
triangle inequality (\ref{eq:Trianle_ineq}), the $2^{\textrm{nd}}$
one is a consequence of the LHS of (\ref{eq:Trianle_ineq}), whereas
the $3^{\textrm{rd}}$ inequality is a bound for the entropy for two-dimensional
Hilbert space of the one-qubit subsystem $\textrm{X}$. Alternatively,
instead of the above reasoning in the part referring to the $1^{\textrm{st}}$
and $2^{\textrm{nd}}$ inequality in (\ref{eq:eqDeltaSIneq}) one
can ignore the RHS of (\ref{eq:Trianle_ineq}) and using arguments
of elementary geometry permute the superscripts in (\ref{eq:Trianle_ineq})
according to $\left(\begin{array}{ccc}
{\scriptstyle \left(\textrm{B}\right)} & {\scriptstyle \left(\textrm{X}\right)} & {\scriptstyle \left(\textrm{BX}\right)}\\
{\scriptstyle \left(\textrm{BX}\right)} & {\scriptstyle \left(\textrm{B}\right)} & {\scriptstyle \left(\textrm{X}\right)}
\end{array}\right)$.

\begin{figure}
\begin{centering}
\begin{tabular}{>{\centering}m{0.45\linewidth}c}
\subfloat[]{\begin{centering}
\includegraphics[scale=0.4]{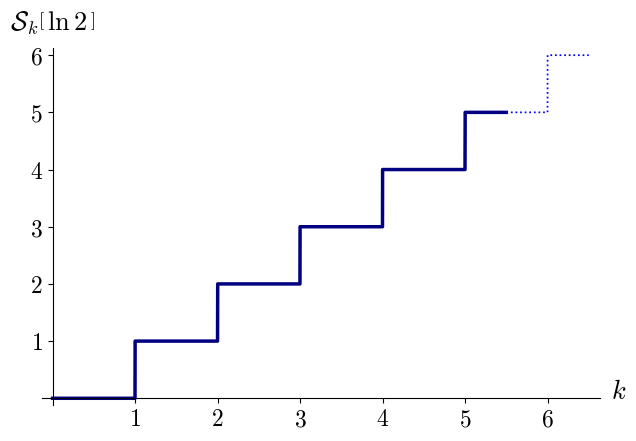}
\par\end{centering}
\label{fig:AEntropyBounds}} & %
\begin{tabular}{>{\centering}p{0.45\linewidth}}
\subfloat[]{\begin{centering}
\includegraphics[scale=0.3]{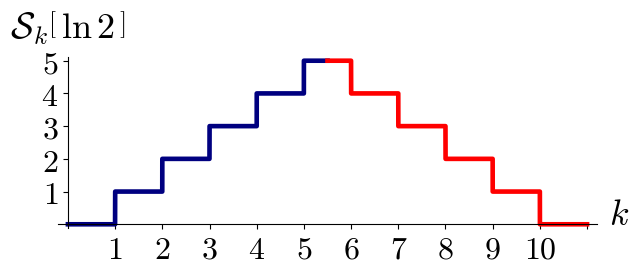}
\par\end{centering}
\label{fig:BEntropyBounds}}\tabularnewline
\subfloat[]{\begin{centering}
\hspace{5mm}\includegraphics[viewport=0mm 0mm 459bp 223bp,scale=0.3]{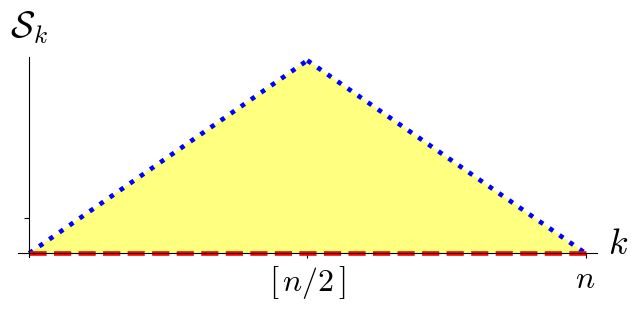}
\par\end{centering}
\label{fig:CEntropyBounds}}\tabularnewline
\end{tabular}\tabularnewline
\end{tabular}
\par\end{centering}
\caption{(Color online) Bounds on entropy curves for unitarily evolving systems,
which follow from semicausality. In particular: (a) presents the (beginning
part of the) curve of the quickest possible growth of entanglement
entropy according to the limitation imposed by (\ref{eq:DoubleSIneq});
(b) takes into account ``symmetric'' limitations (\ref{eq:FinalSIneq})
(here $n=11$); in the limit of a very large $n$, the (shaded) triangle
in (c) presents an allowed area for any entropy curve for semicausal
unitary evolution; here the two upper (dotted) edges of the triangle
form a Page-like curve.}
\label{fig:EntropyBounds}
\end{figure}

Translating (\ref{eq:DeltaSIneq}) onto ``the indexed notation''
of (\ref{eq:EntropySteps}) (see Figure \ref{fig:between2U}), for
each ``time'' step $k$ ($k=1,\ldots,n$), we get the inequality
\begin{equation}
\left|\Delta\mathcal{S}_{k}\right|\equiv\left|\mathcal{S}_{k}^{+}-\mathcal{S}_{k}^{-}\right|\leq\ln2.\label{eq:IndeqDeltaSIneq}
\end{equation}
The description of the evolution of entropy can be simplified a bit
by the fact that entropy is invariant (isometric invariance) with
respect to arbitrary unitary transformations (similarity transformations,
in general), denoted below by the symbol $\mathsf{U}^{(a)}$ and defined
by the operator $U^{\left(a\right)}$ ($a=1,2$), of the corresponding
density matrix, i.e.,
\begin{equation}
\mathsf{U}^{(a)}\left[\mathcal{S}\left(\rho_{k}^{(a)\mp}\right)\right]\equiv\mathcal{S}\left[U^{\left(a\right)}\rho_{k}^{(a)\mp}\left(U^{\left(a\right)}\right)^{-1}\right]=\mathcal{S}\left(\rho_{k}^{(a)\mp}\right).\label{eq:UnitaryInv}
\end{equation}
Therefore, in the series (\ref{eq:EntropySteps}), we can make the
following identifications
\begin{equation}
\mathcal{S}_{k}^{+}\equiv\mathcal{S}\left(\rho_{k}^{\left(a\right)+}\right)=\mathsf{U}_{k}^{\left(a\right)}\mathcal{S}\left(\rho_{k}^{(a)+}\right)=\mathcal{S}\left(\rho_{k+1}^{(a)-}\right)\equiv\mathcal{S}_{k+1}^{-}\equiv\mathcal{S}_{k},\qquad a=1,2,\label{eq:UniInvInStep}
\end{equation}
where the first equality follows from \ref{eq:UnitaryInv} and the
second one follows from the fact that the unitary transformation $\mathsf{U}_{k}^{(a)}$
corresponds to the time evolution (shift in indices) operator $U_{k}^{(a)}$
(entropy is a constant of motion), i.e., in terms of indices alone,
$\mathsf{U}_{k}^{(a)}:\,_{k}^{(a)+}\longmapsto\,_{k+1}^{(a)-}$. In
consequence, (\ref{eq:UniInvInStep}) allows us to a bit simplify
our notation by skipping the superscripts ``$+$'' and ``$-$''
as indicated in the last identity in (\ref{eq:UniInvInStep}). By
identifying neighbouring terms according to (\ref{eq:UniInvInStep}),
the series (\ref{eq:EntropySteps}) can be substantially simplified
and rearranged as follows
\begin{equation}
\mathcal{S}_{0}\left(=0\right)\rightarrow\mathcal{S}_{1}\rightarrow\cdots\rightarrow\mathcal{S}_{k}\rightarrow\cdots\rightarrow\mathcal{S}_{n-1}\rightarrow\mathcal{S}_{n}\left(=0\right).\label{eq:EntStepsRe}
\end{equation}
Now also (\ref{eq:IndeqDeltaSIneq}) assumes a simpler form
\begin{equation}
\left|\Delta\mathcal{S}_{k}\right|=\left|\mathcal{S}_{k}-\mathcal{S}_{k-1}\right|\leq\ln2,\label{eq:ReIndeqDeltaSIneq}
\end{equation}
where $k=1,\ldots,n$. Iteration of (\ref{eq:ReIndeqDeltaSIneq})
yields the solution
\begin{equation}
\mathcal{S}_{k}\leq k\ln2+\mathcal{S}_{0},\label{eq:IterEnIneq}
\end{equation}
or (taking into account the initial condition $\mathcal{S}_{0}=0$)
\begin{equation}
\mathcal{S}_{k}\leq k\ln2,\label{eq:SingSIneq}
\end{equation}
or (by virtue of the non-negativity of entropy)
\begin{equation}
0\leq\mathcal{S}_{k}\leq k\ln2\label{eq:DoubleSIneq}
\end{equation}
(see Figure \ref{fig:AEntropyBounds}).
In the next step, let us utilize (\ref{eq:ReIndeqDeltaSIneq}) once
more. First of all, we can observe that successively inserting the
indices $k=1$, $k=2$, \ldots{} , $k=n-1$, $k=n$ to (\ref{eq:ReIndeqDeltaSIneq})
we obtain the sequence of inequalities $\left|\mathcal{S}_{1}-0\right|\leq\ln2$,
$\left|\mathcal{S}_{2}-\mathcal{S}_{1}\right|\leq\ln2$, \ldots{} ,
$\left|\mathcal{S}_{n-1}-\mathcal{S}_{n-2}\right|\leq\ln2$, $\left|0-\mathcal{S}_{n-1}\right|\leq\ln2$.
Solving that sequence of the inequalities, e.g.\ simply by successive
insertions in the order written (the order of the increasing ``time''
$k$), we obtain (\ref{eq:DoubleSIneq}), but solving it in reversed
order (decreasing ``time'' $k$) yields the inequality

\begin{equation}
0\leq\mathcal{S}_{k}\leq\left(n-k\right)\ln2.\label{eq:RevDoubleSIneq}
\end{equation}
Since the incorporation of semicausality enforces an order in time
evolution, one could wonder whether we are allowed to change the order
of ``time'' $k$. Fortunately, the ``time'' $k$-inversion amounts
to only reversing the order of successive insertions in the process
of solving the above set of inequalities, and the semicausality is
not at risk. Intuitively, it could also be justified by the observation
that even though the theoretical input we use is the notion of semicausality,
all diagrams in Figures 1, 2, 3, and 4 look time-symmetric. Alternatively,
purely formally, making use of the fact that the time-inversed evolution
$U^{\intercal}$ (where the symbol ``$^{\intercal}$'' denotes time-inversion)
of the given time evolution $U$ is also time evolution, we could
repeat our analysis for the new (time-inversed) evolution $U^{\intercal}=U^{-1}$
obtaining (\ref{eq:RevDoubleSIneq}). The logical conjunction of
(\ref{eq:DoubleSIneq}) and (\ref{eq:RevDoubleSIneq}) yields our
final (double) inequality
\begin{equation}
0\leq\mathcal{S}_{k}\leq\min\left(k,n-k\right)\ln2,\quad k=0,1,\ldots,n,\label{eq:FinalSIneq}
\end{equation}
which is illustrated (for $n=11$) by Figure \ref{fig:BEntropyBounds}.

\subsubsection{\label{subsec:The-entropy-bound}The entropy curve for (semi)causal
evolution}

For huge $n$ according to (\ref{eq:FinalSIneq}) the lower and the
upper bounds for the entropy $\mathcal{S}_{k}$ can be graphically
represented as the lower (dashed) edge and the two upper (dotted)
edges of the (shaded) triangle in Figure \ref{fig:CEntropyBounds},
respectively. In other words, for any unitary semicausal evolution,
the entropy curve $\mathcal{S}_{k}$ is supposed to satisfy the following\emph{
semicausal entropy bound(s)}:
\begin{enumerate}
\item The curve $\mathscr{C}$, a graphic representation of the function
$\mathcal{S}_{k}$, should lie inside the shaded triangle ($\vartriangle$)
depicted in Figure \ref{fig:CEntropyBounds} ($\mathscr{C}\subseteq\vartriangle$).
\item Its slope (derivative, in an appropriate continuous limit) should
be bounded by the inequality 
\begin{equation}
\left|\dfrac{d\mathcal{S}}{dk}\right|\leq\ln2.\label{eq:EnBound}
\end{equation}
 (\ref{eq:EnBound}) is a differential version of (\ref{eq:ReIndeqDeltaSIneq}).
\end{enumerate}

\section{\label{sec:Some-examples-of}Examples of semicausal models}

To illustrate ideas concerning the (semi)causal structure of unitary
models of BH evolution (evaporation) introduced in the previous section,
we present two large multiparametric families of qubit toy models:
(1) a family of tensor product models defined in Subsection \ref{subsec:Product-models};
(2) a family of controlled non-product models introduced in Subsection
\ref{subsec:Non-product-models}. In Subsubsection \ref{subsec:Three-qubit-model}
for pedagogical reasons, we start from a very simple three-qubit (3Q)
model, which can serve as a building block for subsequent constructions.
In Subsubsection \ref{subsec:Tensor-square-of}, we ``tensor exponentiate''
the 3Q model, first yielding its tensor square ($3\textrm{Q}^{2}$),
and next, in Subsubsection \ref{subsec:Arbitrary-tensor-power}, its
$m$-th tensor power ($3\textrm{Q}^{m}$). In Subsection \ref{subsec:Non-product-models}
as a less straightforward generalization of the $3\textrm{Q}$ model,
we propose a family of non-product models. In particular, Subsubsection
\ref{subsec:Controlled-three-qubit-model} introduces a four-qubit
controlled version of the 3Q model, which we can call the controlled
three-qubit (C3Q) model. A (non-product) generalization of the C3Q
model to an arbitrarily large number $n$ of qubits is proposed in
Subsubsection \ref{subsec:Controlled-multiqubit-models}. Obviously,
by virtue of the construction, all the introduced models are unitary
and semicausal, and in consequence, they satisfy the entropy bounds
of Subsection \ref{subsec:Bounds-on-entanglement}.

Since all the proposed models of BH evaporation are formulated in
a rather formal language of qubits and quantum circuits, some explanations
in terms of physics should be given. To begin with, let us observe
that there is a direct correspondence between qubit models and two-state
spin models. In turn, at least in principle, two-state spin models
can be translated (e.g., via the Jordan\textendash Wigner transformation)
onto fermion models. Keeping that observation in mind, for our discussion,
we can loosely, i.e., ignoring the actual Jordan\textendash Wigner
transformation, assume the following convention: the qubit $\left|0\right\rangle $
can be associated to a vacuum state, whereas the qubit $\left|1\right\rangle $
can be associated to a one-particle state.

In physics terms, the picture of the BH evolution depicted in Figure
\ref{fig:iterSchumacherSym} looks as follows. The vertical dashed
left line and the right line denote the theoretically earliest moment
and the latest one, respectively, when the BH is allowed to exist\textemdash no
causal restrictions (single vertical boxes $U_{0}$ and $U_{n}$)
and then no event horizons. Obviously, in more realistic many-particle
situations, the BH forms much later (because usually, a huge number
of particles is necessary for the actual BH creation) and decays earlier
(it seems reasonable to assume that a too-low number of particles
is unable to support the existence of the BH). Anyway, when a state
(for some $k$ close to $n$, i.e., for the final stages of BH evolution)
quitting the upper box $U_{k-1}^{(1)}$ (potentially belonging to
a BH) is the vacuum state $\left(\ket{\psi_{k}}^{(1)}=\ket{0}^{(1)}\right)$,
i.e., the total state $\ket{\psi_{k}}$ assumes the particular (product)
form $\ket{\psi_{k}}=\ket{0}^{(1)}\otimes\ket{\psi_{k}}^{(2)}$, certainly,
the BH already (at this stage $k$) does not exist (has evaporated)\textemdash the
vacuum state $\ket{0}^{(1)}$ cannot support the existence of the
BH. Consequently, the rest of the unitary operators (corresponding
to $i=k,k+1,\ldots,n-1$) ceases to be essential from the point of
view of the BH evolution (evaporation) and thus can be put in the
trivial form $U_{i}=\mathbb{I}$ or (equivalently) can be simply skipped.

\subsection{\label{subsec:Product-models}Product models}

The product models discussed in this subsection are tensor products
($m$-th tensor powers) of the 3Q model defined in Subsubsection \ref{subsec:Three-qubit-model}.
Obviously, $m$-th tensor powers are algebraically unique for a given
$m$, but as will be observed in Subsubsection \ref{subsec:Tensor-square-of},
as far as the (semi)causal structure is concerned, there is a multitude
of different product models for each $m$. From combinatorics, it
follows that there are precisely $2^{-m}\left(2m\right)!$ models
for a given $m$, which in general can yield different entropy curves.

\subsubsection{\label{subsec:Three-qubit-model}Three-qubit model}

The three-qubit (3Q) model first appeared (rather sketchy and in ``a
four-qubit incarnation'') in the context of firewalls in \citep{Almheiri2013}
and then directly and explicitly in \citep{Osuga2018}. Another but
akin version of the 3Q model was proposed and discussed in \citep{Broda2020}
(see also \citep{Broda2021}), whereas its some four-qubit extension
in \citep{Broda2021PLB}. In this Subsubsection, the 3Q model appears
for pedagogical reasons as well as an elementary building block for
our further constructions. Our present approach closely follows \citep{Broda2020}.
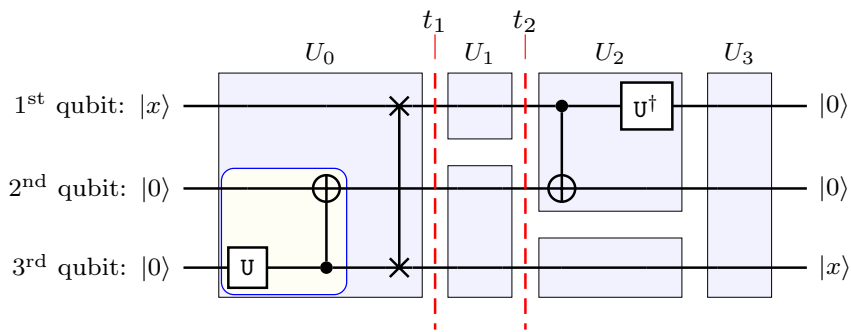
\begin{figure}[h]
\begin{centering}
\def\gg#1#2{\gategroup[{#1},steps={#2},
style={very thin,fill=blue!5,inner sep=0em},background]}
\def\ggbis#1#2{\gategroup[{#1},steps={#2},
style={thin,rounded corners,color=blue,fill=yellow!5,inner sep=-.1em}, background]}
\def\md{{\color{red}\mid}}
\begin{quantikz}
\ls{$1^{\textrm{st}}\textrm{\ qubit:\ }\ket{x}$} &\qw\gg{3}{3}{$U_0$}   &\qw       &\swap{2}\se{$\overset{\textrm{\normalsize$t_1$}}{\md}$} &\qw\gg{1}{2}{$U_1$}&\qw\se{$\overset{\textrm{\normalsize$t_2$}}{\md}$} &\ctrl{1}\gg{2}{2}{$U_2$} &\gt{\mathtt{U}^{\dagger}} &\qw\gg{3}{2}{$U_3$}&\qw&\rs{$\ket{0}$}\qw \\
\ls{$2^{\textrm{nd}}\textrm{\ qubit:\ }\ket{0}$} &\qw\ggbis{2}{2}{}   &\targ{}   &\qw      &\qw\gg{2}{2}{}&\qw &\targ{}  &\qw   &\qw&\qw&\rs{$\ket{0}$}\qw \\
\ls{$3^{\textrm{rd}}\textrm{\ qubit:\ }\ket{0}$} &\gt{\mathtt{U}} &\ctrl{-1} &\targX{} &\qw&\qw &\qw \gg{1}{2}{}     &\qw   &\qw&\qw&\rs{$\ket{x}$}\qw\\[-2ex]
&&&&&&&&&&
\end{quantikz}
\par\end{centering}
\caption{(Color online) Quantum circuit defining the three-qubit (3Q) model.
The first qubit, in the state $\ket{x}$, is interpreted as a ``matter''
qubit entering the BH, whereas the other two (vacuum) qubits are utilized
for Hawking particle production. The Hawking effect parametrized by
$\mathtt{U}$ takes place in the rounded box (see (\ref{eq:HGate}))
in the left lower corner of the circuit.}
\label{fig:T-QModel}
\end{figure}

The 3Q model can be concisely described in terms of a quantum circuit,
as presented in Figure \ref{fig:T-QModel}. In the notation of (\ref{eq:U(many_x)_tensored_generalized})
(see Figure \ref{fig:iterSchumacherSym}), the unitary evolution $U$
of the 3Q model assumes the following block structure
\begin{equation}
U=U_{3}U_{2}U_{1}U_{0}\equiv U_{3}\left(U_{2}^{\left(1\right)}\otimes U_{2}^{\left(2\right)}\right)\left(U_{1}^{\left(1\right)}\otimes U_{1}^{\left(2\right)}\right)U_{0}.\label{eq:UforThreeQM}
\end{equation}
Here the first unitary block $U_{0}$ implements the Hawking process
on the second and third qubit, followed by a swap operation on the
first and third one, $U_{1}=U_{3}=\mathbb{I}$ (identity), and $U_{2}=U_{2}^{(1)}\otimes U_{2}^{(2)}$,
where $U_{2}^{(1)}$ can be interpreted as an inverse of the Hawking
process and $U_{2}^{(2)}=\mathbb{I}$.

\begin{figure}
\subfloat[]{\begin{centering}
\includegraphics[scale=0.25]{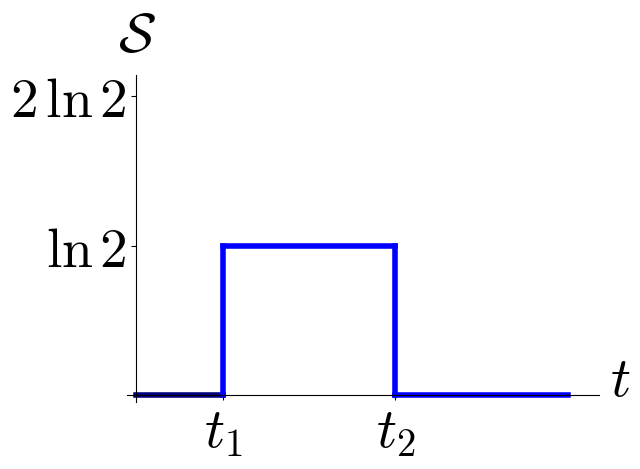}
\par\end{centering}
\label{fig:AEntCur}}\hspace*{\fill}\subfloat[]{\begin{centering}
\includegraphics[scale=0.25]{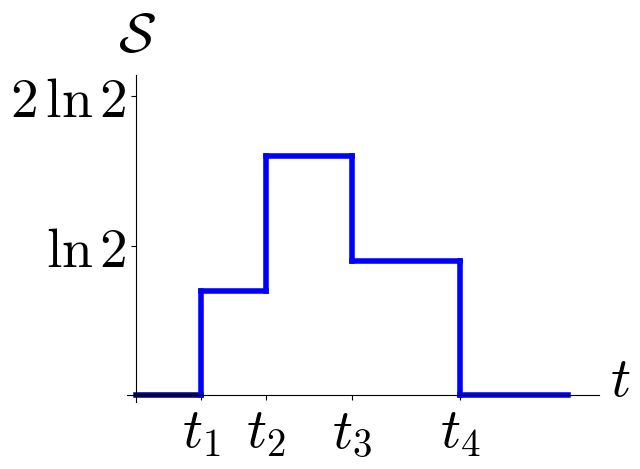}
\par\end{centering}
\label{fig:BEntCur}}\hspace*{\fill}\subfloat[]{\begin{centering}
\includegraphics[scale=0.25]{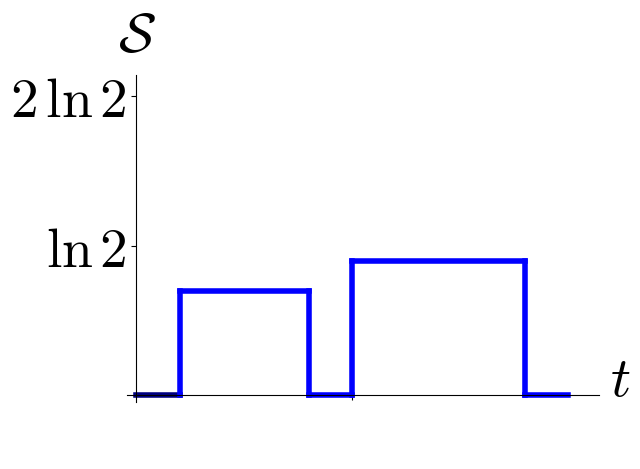}
\par\end{centering}
\label{fig:CEntCur}}

\caption{(Color online) Example entropy curves for the simplest few-qubit product
models. (a) Possible evolution of entanglement entropy for the three-qubit
(3Q) model. In this particular example, in the time interval between
$t_{1}$ and $t_{2}$, the entropy attains its (maximum) value $S=\ln2$
($t_{1}$ and $t_{2}$ are defined in Figure \ref{fig:T-QModel}).
In general, i.e., for arbitrary $\mathtt{U}$, $0\protect\leq S\protect\leq\ln2$.
(b) Possible evolution of entanglement entropy for the tensor square
of the three-qubit ($3\textrm{Q}^{2}$) model. The entropy curve belongs
to ``the Page type'' (the cases (1)\textendash (4) on the list (\ref{eq:SixPoss})).
In particular, for $S_{1}=S_{2}=\ln2$ (the extremal case), we would
exactly obtain ``the two-qubit Page curve''. $t_{1}$, $t_{2}$,
$t_{3}$, and $t_{4}$ are defined in Figure \ref{fig:TST-QModel}.
(c) Another type of possible evolution of entanglement entropy for
the $3\textrm{Q}^{2}$ model. This time the curve is ``minimal''
rather than of ``the Page type''\textemdash the cases (5)\textendash (6)
on the list (\ref{eq:SixPoss}).}
\label{fig:EntCur}
\end{figure}

In principle, the one-qubit operator $\mathtt{U}$ in Figure \ref{fig:T-QModel}
could be an arbitrary unitary operator, but we can as well assume
that it is orthogonal, given in the form
\begin{equation}
\mathtt{U}=\mathtt{U}_{r_{p}}=\left(\begin{array}{cc}
\cos r_{p} & -\sin r_{p}\\
\sin r_{p} & \cos r_{p}
\end{array}\right),\label{eq:defU}
\end{equation}
because of the correspondence to the Hawking process for fermions,
described by orthogonal (Bogolyubov) transformations. More precisely,
the gate

\begin{equation}
\label{eq:HGate}
\def\gg#1#2{\gategroup[{#1},steps={#2},
style={thin,rounded corners,color=blue,fill=yellow!5,inner sep=-.1em}, background]} 
\begin{quantikz}
\ls{$\ket{0}$}	&\qw\gg{2}{2}{}		 &\targ{}	&\qw\rstick[wires=2]%
{\ $\cos r_{p}\ket{00}+\sin r_{p}\ket{11}$}	\\
\ls{$\ket{0}$}	&\gt{\mathtt{U}_{r_p}}  &\ctrl{-1}  &\qw	
\end{quantikz}
\end{equation}\\
implements the (one-mode $p$) Hawking process for fermions with mass
$m$ and momentum $p$, where
\begin{equation}
\cos r_{p}=\left[2\cosh\left(\pi\omega_{p}/\kappa\right)\right]^{-1/2}\exp\left(\pi\omega_{p}/2\kappa\right),\quad\omega_{p}=\sqrt{p^{2}+m^{2}},\label{eq:cosrp}
\end{equation}
with $\kappa$ denoting the surface gravity \citep{Mann2015}.

In the 3Q model, the entanglement entropy $\mathcal{S}$ only changes
at the instants $t_{1}$ (there is a jump of the numeric value of
$\mathcal{S}$ from 0 to some non-negative value $S\leq\ln2$) and
$t_{2}$ (a jump from the value $S$ to $0$ back), as depicted in
Figure \ref{fig:AEntCur} ($t_{1}$ and $t_{2}$ are defined in Figure
\ref{fig:T-QModel}). In terms of the four blocks $U_{0},\ldots,U_{3}$
of Figure \ref{fig:T-QModel}, the evolution of entanglement entropy
proceeds according to the pattern

\begin{equation}
\underset{0}{0}\rightarrow\underset{1}{S}\rightarrow\underset{2}{0}\rightarrow\underset{3}{0},\label{eq:EEntropy43Q}
\end{equation}
where the underscripts number the unitary blocks. The evolution of
entropy can be called extremal, if the entropy attains its maximal
allowed value, i.e., for this model $S=\ln2$. According to (\ref{eq:cosrp})
this happens in the limit of $\kappa\rightarrow\infty$ or $\omega_{p}\left(=\sqrt{p^{2}+m^{2}}\right)\rightarrow0$
(maximal entanglement, see (\ref{eq:HGate})).

\subsubsection{\label{subsec:Tensor-square-of}Tensor square of the three-qubit
model}

As an intermediate stage between the 3Q model and its arbitrary $m$-th
tensor power $\left(3\textrm{Q}^{m}\right)$, we will consider the
$3\textrm{Q}^{2}$ model, which is a tensor square of the 3Q model.
The $3\textrm{Q}^{2}$ model operates on six qubits, and its unitary
evolution $U$ can be semicausally decomposed into $1+6\left(\textrm{number of qubits}\right)=7$
blocks

\begin{equation}
\begin{aligned}U= & U_{6}U_{5}U_{4}U_{3}U_{2}U_{1}U_{0}\\
\equiv & U_{6}\left(U_{5}^{\left(1\right)}\otimes U_{n-1}^{\left(2\right)}\right)\ldots\left(U_{2}^{\left(1\right)}\otimes U_{2}^{\left(2\right)}\right)\left(U_{1}^{\left(1\right)}\otimes U_{1}^{\left(2\right)}\right)U_{0}.
\end{aligned}
\label{eq:UforTSThreeQM}
\end{equation}

\begin{figure}[h]
\begin{centering}
\def\sw{\swap{4}}
\def\tx{\targX{}}
\def\tg{\targ{}}
\def\cu{\ctrl{-2}}
\def\cd{\ctrl{2}}
\def\ls#1{\lstick{$\ket{{#1}}$}}
\def\rs#1{\rstick{$\ket{{#1}}$}}   
\def\gg#1#2{\gategroup[{#1},steps={#2},
style={very thin,fill=blue!5,inner sep=.15em, outer sep=0em},background]}
\def\md{{\color{red}\mid}}
\def\gt#1{\gate{\mathtt{U}_{#1}}}
\def\gd#1{\gate{\mathtt{U}_{#1}^\dagger}}
\begin{quantikz}
\ls{x_1}&\qw\gg{6}{3}{$U_0$}&\qw&\sw\se{$\overset{\textrm{\normalsize$t_1$}}{\md}$} &\qw\gg{1}{2}{$U_1$}&
\qw\se{$\overset{\textrm{\normalsize$t_2$}}{\md}$} 
&[.5em]
\qw\gg{2}{1}{$U_2$}\se{$\overset{\textrm{\normalsize$t_3$}}{\md}$}
&[.5em]
\cd\gg{3}{1}{$U_3$}\se{$\overset{\textrm{\normalsize$t_4$}}{\md}$} &\gd{1}\gg{4}{1}{$U_4$} &\qw\gg{5}{1}{$U_5$} &\qw\gg{6}{1}{$U_6$}&\rs{0}\qw	\\
\ls{x_2}&\qw&\qw&\qw &\qw\gg{5}{2}{}&\sw&\qw &\qw &\cd &\gd{2} &\qw&\rs{0}\qw	\\
\ls{0}&\qw&\tg&\qw &\qw&\qw &\qw\gg{4}{1}{} &\tg &\qw &\qw &\qw&\rs{0}\qw	\\
\ls{0}&\qw&\qw&\qw &\tg&\qw &\qw &\qw\gg{3}{1}{} &\tg &\qw &\qw&\rs{0}\qw	\\
\ls{0}&\gt{1}&\cu&\tx &\qw&\qw &\qw &\qw &\qw\gg{2}{1}{} &\qw &\qw&\rs{x_1}\qw	\\
\ls{0}&\qw&\qw&\gt{2} &\cu&\tx &\qw &\qw &\qw &\qw\gg{1}{1}{} &\qw&\rs{x_2}\qw
\end{quantikz}
\par\end{centering}
\caption{(Color online) Quantum circuit defining one of the 6 variants of the
$3\textrm{Q}^{2}$ model (the first case on the list of possible semicausal
embeddings (\ref{eq:SixPoss})). The first two qubits, in the states
$\ket{x_{1}}$ and $\ket{x_{2}}$, are interpreted as ``matter''
qubits entering the BH, whereas the other four (vacuum) qubits are
utilized for Hawking particle production.}
\label{fig:TST-QModel}
\end{figure}
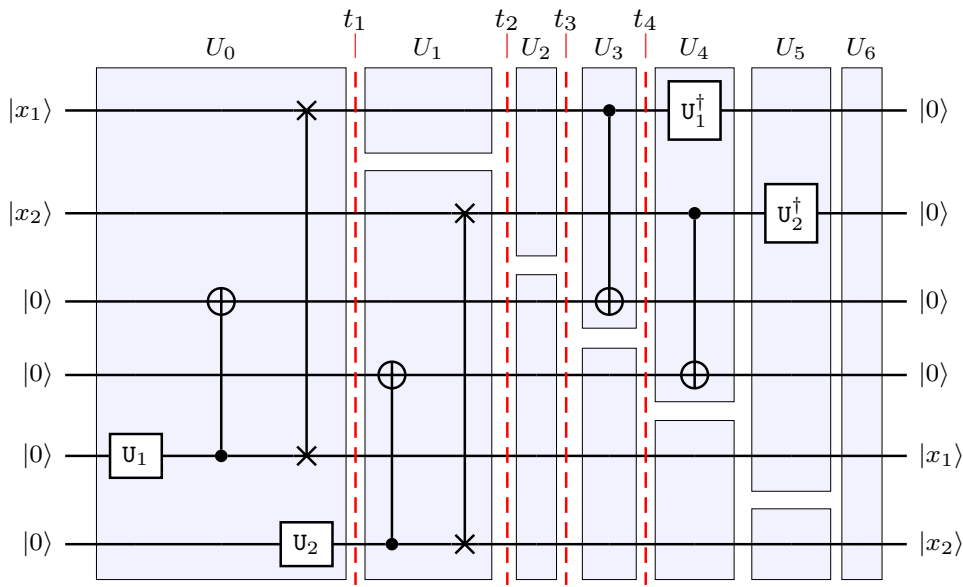

\begin{figure}
\begin{centering}
\hspace*{\fill}\subfloat[]{\begin{centering}
\def\gg{\gategroup[2,steps=1,style={dotted,rounded corners,color=green,fill=magenta!5,inner sep=0ex},background]}
\begin{quantikz}
\begin{array}{c}1^{\textrm{st}}\textrm{ triple:}\\
\ket{x_1},\ket{0},\ket{0}\end{array}\qquad
&\gate{\begin{array}{c} \textrm{Figure \ref{fig:T-QModel}} \\
\textrm{with}\\ \mathtt{U}=\mathtt{U}_1 \end{array}}%
\qwbundle[alternate]{}\gg{}
&\qwbundle[alternate]{}\gg{}
&\qwbundle[alternate]{}
\\[-2.5ex]
\begin{array}{c}2^{\textrm{nd}}\textrm{ triple:}\\
\ket{x_2},\ket{0},\ket{0}\end{array}\qquad
&\qwbundle[alternate]{}
&\gate{\begin{array}{c} \textrm{Figure \ref{fig:T-QModel}} \\
\textrm{with}\\ \mathtt{U}=\mathtt{U}_2 \end{array}}%
\qwbundle[alternate]{}
&\qwbundle[alternate]{}
\end{quantikz}
\par\end{centering}
\label{fig:ATST-QModel}}\hspace*{\fill}\subfloat[]{\centering{}\def\gg{\gategroup[2,steps=1,style={dotted,rounded corners,color=green,fill=magenta!5,inner sep=0ex},background]}
\begin{quantikz}
%
&\qwbundle[alternate]{}\gg{}
&\gate{\begin{array}{c} \textrm{Figure \ref{fig:T-QModel}} \\
\textrm{with}\\ \mathtt{U}=\mathtt{U}_1 \end{array}}\qwbundle[alternate]{}\gg{}
&\qwbundle[alternate]{}
\\[-2.5ex]
&\gate{\begin{array}{c} \textrm{Figure \ref{fig:T-QModel}} \\
\textrm{with}\\ \mathtt{U}=\mathtt{U}_2 \end{array}}\qwbundle[alternate]{}
&\qwbundle[alternate]{}
&\qwbundle[alternate]{}
\end{quantikz}\label{fig:BTST-QModel}}\hspace*{\fill}
\par\end{centering}
\centering{}\caption{(Color online) Schematic presentation of quantum circuits defining
the last two cases on the list (\ref{eq:SixPoss}) of all possible
semicausal embeddings of the $3\textrm{Q}^{2}$ model. The cases (5)
and (6), depicted in (a) and (b),  respectively, belong to the ``minimal''
(or ``decoupled'') type, i.e., the shape of their entropy curves
is similar to the shape of the curve depicted in Figure \ref{fig:CEntCur}.}
\label{fig:2TST-QModels}
\end{figure}
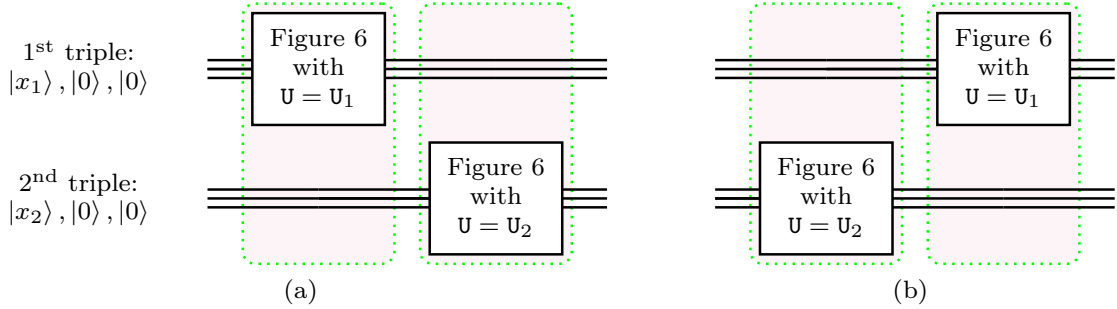

One of the versions of the model (a preferred one from our perspective)
is explicitly presented in Figure \ref{fig:TST-QModel}.
As already mentioned at the beginning of Subsection \ref{subsec:Product-models},
even though from a purely algebraic point of view the tensor square
(and more generally, the tensor power) is unique, the corresponding
models are not, because of various possibilities of embeddings in
the surrounding semicausal frame (Figure \ref{fig:iterSchumacherSym}).
That follows from the fact that particular qubits can enter successive
upper subboxes $\left(U_{1}^{(1)},U_{2}^{(1)},\ldots,U_{n-1}^{(1)},U_{n}\right)$
in various orders, giving rise to (in principle) different entropy
curves. For the $3\textrm{Q}^{2}$ model, we have six possibilities,
which can be identified by the following entropy jumps:

\begin{equation}
\begin{alignedat}{1}(1)\qquad & \underset{0}{0}\rightarrow\underset{1}{S_{1}}\rightarrow\underset{2}{S_{1}+S_{2}}\rightarrow\underset{3}{S_{2}}\rightarrow\underset{4}{0}\rightarrow\underset{5}{0}\rightarrow\underset{6}{0};\\
(2)\qquad & \underset{}{0}\rightarrow\underset{}{S_{1}}\rightarrow\underset{}{S_{1}+S_{2}}\rightarrow\underset{}{S_{1}}\rightarrow\underset{}{0};\\
(3)\qquad & \underset{}{0}\rightarrow\underset{}{S_{2}}\rightarrow\underset{}{S_{1}+S_{2}}\rightarrow\underset{}{S_{1}}\rightarrow\underset{}{0};\\
(4)\qquad & \underset{}{0}\rightarrow\underset{}{S_{2}}\rightarrow\underset{}{S_{1}+S_{2}}\rightarrow\underset{}{S_{2}}\rightarrow\underset{}{0};\\
(5)\qquad & \underset{}{0}\rightarrow\underset{}{S_{1}}\rightarrow\underset{}{0}\rightarrow\underset{}{S_{2}}\rightarrow\underset{}{0};\\
(6)\qquad & \underset{}{0}\rightarrow\underset{}{S_{2}}\rightarrow\underset{}{0}\rightarrow\underset{}{S_{1}}\rightarrow\underset{}{0.}
\end{alignedat}
\label{eq:SixPoss}
\end{equation}
The first case (1) on the list (\ref{eq:SixPoss}) (with explicitly
numbered blocks $U_{0},\ldots,U_{6}$ by the underscripts $0,1.\ldots,6$)
corresponds to the model presented in Figure \ref{fig:TST-QModel},
and its entropy jumps are qualitatively illustrated in Figure \ref{fig:BEntCur};
the second case (2) is some variation of (1); the cases (3) and (4)
are ``qubit transpositions'' of (1) and (2), respectively; whereas
the cases (5) and (6) are ``causally decoupled'' models. The cases
(1)\textendash (4) belong to ``the Page type'', i.e., the shape
of their entropy curves resembles the shape of the curve depicted
in Figure \ref{fig:BEntCur}; in particular, for $S_{1,2}=\ln2$ (the
extremal case), we exactly obtain ``the two-qubit Page curve''.
The cases (5) and (6) schematically presented in Figure \ref{fig:ATST-QModel}
and Figure \ref{fig:BTST-QModel}, respectively, belong to the ``minimal''
type, i.e., the shape of their entropy curves resembles the shape
of the curve depicted in Figure \ref{fig:CEntCur}.

One should also note that $S_{1}$ and $S_{2}$ $\left(0\leq S_{1,2}\leq\ln2\right)$
are contributions to the entanglement entropy $\mathcal{S}$ coming
from the Hawking effect rather than from the ``matter'' qubits $\ket{x_{1}}$
and $\ket{x_{2}}$.

\subsubsection{\label{subsec:Arbitrary-tensor-power}Arbitrary tensor power of the
three-qubit model}

From the point of view of entropy curves (or entropy evolution), various
$3\textrm{Q}^{m}$ models with the fixed tensor power $m$ can be
characterized by various series of entropy contributions $\pm S_{1},\pm S_{2},\ldots,\pm S_{m}$,
where the numeric values $S_{i}$ ($0\leq S_{i}\leq\ln2$, $i=1,2,\ldots,m$)
of the entanglement entropy $\mathcal{S}$ depend on the explicit
form of the corresponding $U(2)$ matrices $\mathtt{U}_{i}$. For
example, the subset of all $3\textrm{Q}^{m}$ models corresponding
to entropy jumps (analogous to the cases (1)\textendash (4) on the
list (\ref{eq:SixPoss}))
\begin{equation}
\begin{alignedat}{1}0\rightarrow & S_{\sigma_{+}\left(1\right)}\rightarrow S_{\sigma_{+}\left(1\right)}+S_{\sigma_{+}\left(2\right)}\rightarrow\cdots\rightarrow S_{\sigma_{+}\left(1\right)}+S_{\sigma_{+}\left(2\right)}+\cdots+S_{\sigma_{+}\left(m\right)}\\
\rightarrow & S_{\sigma_{+}\left(1\right)}+S_{\sigma_{+}\left(2\right)}+\cdots+S_{\sigma_{+}\left(m\right)}-S_{\sigma_{-}\left(1\right)}\\
\rightarrow & S_{\sigma_{+}\left(1\right)}+S_{\sigma_{+}\left(2\right)}+\cdots+S_{\sigma_{+}\left(m\right)}-S_{\sigma_{-}\left(1\right)}-S_{\sigma_{-}\left(2\right)}\rightarrow\cdots\rightarrow0,
\end{alignedat}
\label{eq:PagePoss}
\end{equation}
where $\sigma_{+}$ and $\sigma_{-}$ are two independent permutations
of $m$ elements, is of ``the Page type'', i.e., for $S_{i}\approx\ln2$
the shape of the entropy curves for these models is similar to the
shape of the curve depicted in Figure \ref{fig:BEntropyBounds} or
Figure \ref{fig:CEntropyBounds} for $m\gg1$ (two upper edges of
the triangle). Instead, the subset of $3\textrm{Q}^{m}$ models corresponding
to entropy jumps (analogous to the cases (5)\textendash (6) in (\ref{eq:SixPoss}))
\begin{equation}
0\rightarrow S_{\sigma\left(1\right)}\rightarrow0\rightarrow S_{\sigma\left(2\right)}\rightarrow0\rightarrow\cdots\rightarrow0\rightarrow S_{\sigma\left(m\right)}\rightarrow0,\label{eq:ZeroPoss}
\end{equation}
where $\sigma$ is a (single) permutation of $m$ elements, belongs
to the ``minimal'' type, i.e., for $m\gg1$ their entropy curves
are very close to the lower (dashed) edge of the (shaded) triangle
in Figure \ref{fig:CEntropyBounds}. Let us also note that the number
of models belonging to the Page type (\ref{eq:PagePoss}) is equal
to $\left(m!\right)^{2}$, and for $m\gg1$ this number is much larger
than the number of ``minimal'' models (\ref{eq:ZeroPoss}), which
is equal to $m!$.

\subsection{\label{subsec:Non-product-models}Non-product models}

A lot of (product) models introduced in the previous Subsection \ref{subsec:Product-models}
can yield entropy curves arbitrarily close to the Page curve, but
from the point of view of physics, the models can seem to be almost
trivial\textemdash interactions are bounded to mutually isolated and
almost identical (i.e., possibly differing only by one-qubit operators
$\mathtt{U}_{i}$ or equivalently by the parameters $r_{p_{i}}$ in
(\ref{eq:defU})) three-qubit sectors. The Hawking process in these
models is determined by the fixed family of matrices $\left\{ \mathtt{U}_{i}\right\} _{i=1}^{m}$.
Therefore, the whole system looks like an \emph{open} one defined
by externally prescribed $r_{p_{i}}$, $i=1,2,\ldots,m$. To remove
these limitations, in this subsection we introduce a family of controlled
non-product models, which are controlled with a relatively small set
of $l$ control qubits $\left\{ \left|y_{i}\right\rangle \right\} _{i=1}^{l}$.
The control qubits should be interpreted as the part of all ``matter''
qubits that gradually enter the BH, but their modest number is unable
to support the existence of the BH in the final stages of its evolution.
Thus, the total number of qubits in these models amounts to $n=l+3m$
($l$ control qubits and $3m$ ``controlled'' qubits, including
$m$ ``matter'' qubits as well as $2m$ Hawking ones), where for
large $m$ (the case of BHs), we should assume that $l\ll m$. The
simplest case, $l=1$ and $m=1$, i.e., the controlled three-qubit
(C3Q) model, is discussed in the following Subsubsection \ref{subsec:Controlled-three-qubit-model},
whereas the case of arbitrary $l$ and $m$, i.e., the $l$-controlled
$3m$-qubit ($l\textrm{C}3m\textrm{Q}$) model, is introduced in Subsubsection
\ref{subsec:Controlled-multiqubit-models}.

\subsubsection{\label{subsec:Controlled-three-qubit-model}The controlled three-qubit
model}

The controlled three-qubit (C3Q) model operates on four qubits (see
Figure \ref{fig:C3QModel}):
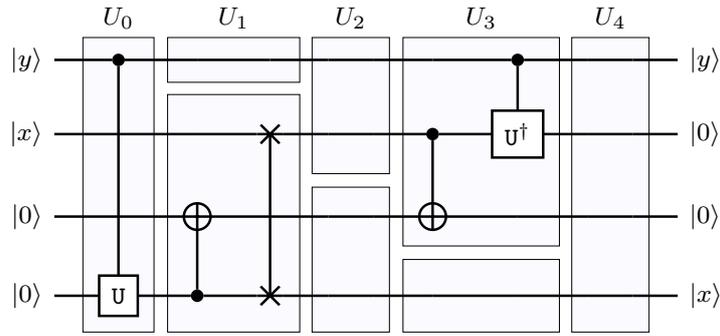
\begin{figure}
\begin{centering}
\def\gg#1#2{\gategroup[{#1},steps={#2},
style={very thin,fill=blue!2,inner sep=.25em},background]}
\begin{quantikz}
\ls{$\ket{y}$}&\ctrl{3}\gg{4}{1}{$U_0$} &\qw\gg{1}{2}{$U_1$} &\qw &\qw\gg{2}{2}{$U_2$} &\qw
&\qw\gg{3}{2}{$U_3$} &\ctrl{1} &\qw\gg{4}{2}{$U_4$} &\qw &\rs{$\ket{y}$}\qw \\
\ls{$\ket{x}$}&\qw &\qw\gg{3}{2}{} &\swap{2} &\qw &\qw &\ctrl{1} 
&\gate{\mathtt{U}^\dagger} &\qw &\qw &\rs{$\ket{0}$}\qw \\
\ls{$\ket{0}$}&\qw &\targ{} &\qw &\qw\gg{2}{2}{} &\qw &\targ{} &\qw &\qw &\qw &\rs{$\ket{0}$}\qw \\
\ls{$\ket{0}$}&\gt{\mathtt{U}} &\ctrl{-1} &\targX{} &\qw &\qw &\qw\gg{1}{2}{} &\qw &\qw &\qw &\rs{$\ket{x}$}\qw
\end{quantikz}
\par\end{centering}
\caption{Quantum circuit defining the controlled three-qubit (C3Q) model. The
first two qubits, in the states $\ket{y}$ (a control qubit) and $\ket{x}$,
are ``matter'' qubits, whereas the other two vacuum states are utilized
for Hawking particle production. In the first unitary block ($U_{0}$)
the qubit $\ket{y}$ controls the Hawking effect by the CNOT-$\mathtt{U}$
gate determined by the $U(2)$ matrix $\mathtt{U}$.}
\label{fig:C3QModel}
\end{figure}
 two ``matter'' qubits in arbitrary states $\ket{y}$ (a control
qubit) and $\ket{x}$ as well as two ``Hawking qubits'', initially
both in the vacuum state $\ket{0}$. The model is parameterized by
a single one-qubit $U(2)$ matrix $\mathtt{U}$, possibly limited
to the orthogonal (Bogolyubov\textendash Hawking) form (\ref{eq:defU}).
The qubit $\ket{y}$ controls (according to the NOT-rule) the Hawking
effect by the CNOT-$\mathtt{U}$ gate in the block $U_{0}$ of Figure
\ref{fig:C3QModel}. 

The evolution of the entanglement entropy $\mathcal{S}$ proceeds
in consecutive blocks $U_{0},\ldots,U_{4}$, according to the pattern
(compare to (\ref{eq:EEntropy43Q}))

\begin{equation}
\underset{0}{0}\rightarrow\underset{1}{S'}\rightarrow\underset{2}{S''}\rightarrow\underset{3}{0}\rightarrow\underset{4}{0},\label{eq:EEntropy4C3Q}
\end{equation}
where the numeric values of $\mathcal{S}$, i.e., $S'$ and $S''$,
with $0\leq S',S''\leq\ln2$, are determined by the qubit $\ket{y}$
and the explicit form of the matrix $\mathtt{U}$. One should remark
that the contributions $S'$ and $S''$ are coming from the Hawking
particles as well as from the control ``matter'' qubit $\ket{y}$.

Let us note that some immediate and natural generalization of the
C3Q model is possible by the replacement of the CNOT-$\mathtt{U}$
gate (defined with a single $U(2)$ matrix $\mathtt{U}$) in the block
$U_{0}$ of Figure \ref{fig:C3QModel} by a gate determined with a
pair of $U(2)$ matrices $\mathtt{U}_{1}$ and $\mathtt{U}_{2}$,
i.e., for the whole two-qubit unitary operator, we have the generalization

\begin{equation}
\begin{pmatrix}\mathbb{I} & 0\\
0 & \mathtt{U}
\end{pmatrix}\longrightarrow\mathcal{U}\equiv\begin{pmatrix}\mathtt{U}_{1} & 0\\
0 & \mathtt{U}_{2}
\end{pmatrix},\label{eq:gen.contr.u-gate}
\end{equation}
where the matrices $\mathtt{U}_{1}$ and $\mathtt{U}_{2}$ correspond
to the control qubits $\ket{0}$ and $\ket{1}$, respectively.

\subsubsection{\label{subsec:Controlled-multiqubit-models}Controlled multiqubit
models}

The $l$-controlled $3m$-qubit ($l\textrm{C}3m\textrm{Q}$) model
is a direct generalization of the C3Q model of the previous subsubsection,
where instead of a single control qubit $\ket{y}$ we have a (small)
set of $l$ control qubits $\ket{y_{1}},\ket{y_{2}},\ldots,\ket{y_{l}},$
and instead of a single triple of the controlled qubits (initially
in the states $\ket{x}$, $\ket{0}$, $\ket{0}$) we have a set of
$m$ triples (initially in the states $\ket{x_{1}},\ket{0},\ket{0},\ldots,\ket{x_{m}},\ket{0},\ket{0}$).
Then, the total number of qubits amounts to $n=l+3m$. Moreover, for
each triple $i$ ($i=1,2,\ldots,m$) there is defined a collection
of the $2^{l}$ $U(2)$ matrices $\left\{ \mathtt{U}_{j}\left(i\right)\right\} _{j=1}^{2^{l}}$
that determine an $l+1$-qubit unitary matrix $\mathcal{U}_{i}$ of
dimension $2^{l+1}\times2^{l+1}$\textemdash a generalization of (\ref{eq:gen.contr.u-gate})
to $l$ control qubits. Explicitly,

\begin{equation}
\mathcal{U}_{i}=\left(\begin{array}{cc}
\begin{array}{cc}
\mathtt{U}_{1}\left(i\right)\\
 & \mathtt{U}_{2}\left(i\right)
\end{array} & \raisebox{-1ex}{\text{\Huge0}}\\
\kern-1em \raisebox{-1ex}{\text{\Huge0}} & \begin{array}{cc}
\ddots\\
 & \mathtt{U}_{2^{l}}\left(i\right)
\end{array}
\end{array}\right),\label{eq:gen.gen.contr.u-gate}
\end{equation}
where the $U(2)$ matrices $\mathtt{U}_{j}\left(i\right)$ operate
in the $i$-th qubit triple and binary presentation of the number
$j$ corresponds to the series of ``control'' digits given by $y_{1}\cdots y_{l}$.

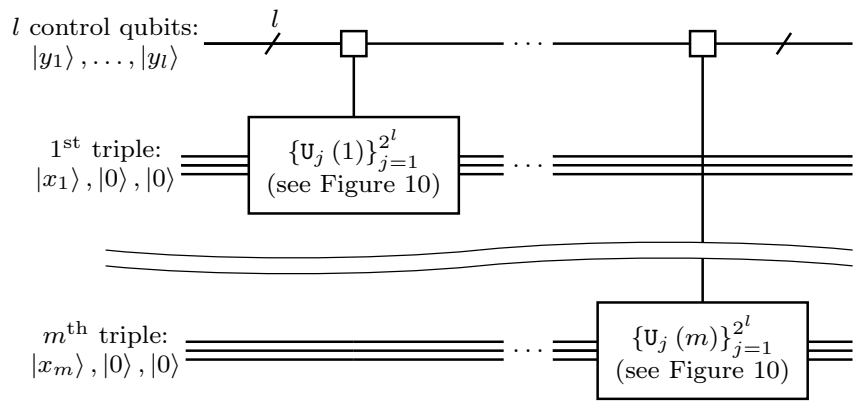
\begin{figure}[h]
\begin{centering}
\def\sp{0em}
\begin{quantikz}
\begin{array}{c}\textrm{$l$ control qubits:}\\
\ket{y_{1}},\ldots,\ket{y_{l}}\end{array}
&[\sp]\gate{}\vqw{1}\qwbundle{\hskip-.5em\raisebox{.5ex}{$l$}} &\ \ldots\ \qw &\gate{}\vqw{3} &\qwbundle{}
\\
\begin{array}{c}1^{\textrm{st}}\textrm{ triple:}\\
\ket{x_1},\ket{0},\ket{0}\end{array}
&\gate{\begin{array}{c} \left\{ \mathtt{U}_{j}\left(1\right)\right\} _{j=1}^{2^{l}} \\ \textrm{(see Figure \ref{fig:C3QModel})}\\
\end{array}}
\qwbundle[alternate]{}  &\ \ldots\ \qwbundle[alternate]{} &\qwbundle[alternate]{}      &\qwbundle[alternate]{}
\\
\wave&&&&
\\
\begin{array}{c}m^{\textrm{th}}\textrm{ triple:}\\
\ket{x_m},\ket{0},\ket{0}\end{array}
&\qwbundle[alternate]{}         
&\ \ldots\ \qwbundle[alternate]{} &\gate{\begin{array}{c} \left\{ \mathtt{U}_{j}\left(m\right)\right\} _{j=1}^{2^{l}} \\ \textrm{(see Figure \ref{fig:C3QModel})}\\
\end{array}}
\qwbundle[alternate]{}  &\qwbundle[alternate]{}
\end{quantikz}
\par\end{centering}
\caption{Schematic presentation (in block form) of the quantum circuit for
the $l$-controlled $3m$-qubit ($l\textrm{C}3m\textrm{Q}$) model
without explicitly imposed semicausal structure. The model operates
on $n\left(=l+3m\right)$ qubits, where for each qubit triple $\left\{ \ket{x_{i}},\ket{0},\ket{0}\right\} $
($i=1,2,\ldots,m$), there is an associated collection of the $2^{l}$
$U(2)$ matrices $\left\{ \mathtt{U}_{j}\left(i\right)\right\} _{j=1}^{2^{l}}$
that define an $l+1$-qubit unitary operator $\mathcal{U}_{i}$ of
the form (\ref{eq:gen.gen.contr.u-gate}). The matrices $\mathtt{U}_{1}\left(i\right)$,
$\mathtt{U}_{2}\left(i\right)$, $\ldots$ , $\mathtt{U}_{2^{l}}\left(i\right)$
act on a single qubit in the box corresponding to the given $i$-th
triple.}
\label{fig:CmQModel}
\end{figure}

The general structure of the $l\textrm{C}3m\textrm{Q}$ model is schematically
depicted in block form in Figure \ref{fig:CmQModel}.
The whole family of the models is parametrized by $m$ $2^{l}$-element
collections of $U(2)$ matrices $\left\{ \mathtt{U}_{j}\left(i\right)\right\} _{j=1}^{2^{l}}$,
$i=1,\ldots,m$, and by a multitude of various semicausal structures.
By virtue of the construction, entropy curves are still within the
semicausal entropy bound given in (\ref{eq:FinalSIneq}) and described
in Subsubsection (\ref{subsec:The-entropy-bound}), but they in a
more complicated way depend on the matrices $\mathtt{U}_{j}\left(i\right)$
as well as on the initial values of the control ``matter'' qubits
$\ket{y_{1}},\ldots,\ket{y_{l}}$.

\subsection{Black hole physics in qubit models}

\noindent Our models incorporate all principal stages of standard
classical and quantum evolution from the point of view of BH physics.
The first era, BH formation, is based on \textquotedbl matter\textquotedbl{}
qubits represented by $\left|x\right\rangle $, $\left|y\right\rangle $,
$\left|x_{i}\right\rangle $ ($i=1,2,\ldots,m$), and $\left|y_{k}\right\rangle $
($k=1,2,\ldots,l$) in the models of Subsubsection \ref{subsec:Three-qubit-model},
\ref{subsec:Controlled-three-qubit-model}, \ref{subsec:Arbitrary-tensor-power},
and \ref{subsec:Controlled-multiqubit-models}, respectively. These
qubits represent all incoming particles that create the BH through
standard classical gravitational collapse. The second era, BH (quantum)
evaporation and disappearance, involves the Hawking process of particle
pair-production. All models include the Hawking mechanism through
gate (\ref{eq:HGate}), which yields the entangled state $\ensuremath{\cos r_{p}\ket{00}+\sin r_{p}\ket{11}}$
and formally implements Hawking pair creation for a fermion mode with
mass $m$ and momentum $p$. The parameters $m$ and $p$ defining
gate (\ref{eq:HGate}) are determined by (\ref{eq:cosrp}) and (\ref{eq:defU}).
We associate qubit $\left|0\right\rangle $ with the vacuum state
and qubit $\left|1\right\rangle $ with a one-particle state, as explained
in the second-to-last paragraph of Section \ref{sec:Some-examples-of}'s
beginning part.

Our models function as toy models, without a specific Hamiltonian
or action. Loosely speaking, the systems represented by our product
models (Subsection \ref{subsec:Product-models}) can be viewed as
qubit analogs of uncoupled oscillators. Non-product models (Subsection
\ref{subsec:Non-product-models}), on the other hand, can be seen
as analogous to specifically coupled oscillators. It is important
to note that there is no widely agreed upon detailed scenario for
the final moments of BH evolution (evaporation), and therefore various
possibilities could be considered. In our product models, the transition
of the state inside either box $U^{\left(X_{1}X_{2}\ldots X_{N-1}\right)}$
in Figure \ref{fig:iterSchumacher} or box $U_{n-1}^{(1)}$ in Figure
\ref{fig:iterSchumacherSym} to the \textquotedbl vacuum\textquotedbl{}
state indicates a peaceful disappearance of the BH. As discussed in
the final paragraph of the paper, in non-product models, we assume
that a small number of control qubits are released only after the
BH has completely ceased to exist, as too few particles are unable
to sustain the BH's existence.

\section{Summary and final remarks}

Utilizing the universal form of semicausal evolution (\ref{eq:U(BXA)_tensored})
(see Figure \ref{fig:Schumacher}), we have established a general
structure of the quantum circuit presenting unitary evolution and
respecting causality imposed by the BH event horizon (see Figures
\ref{fig:iterSchumacherSym} and \ref{fig:between2U}). Moreover,
making use of several standard properties of the von Neumann entropy
(the triangle inequality, isometric invariance, and others), we have
determined principles governing the evolution of the corresponding
BH entanglement entropy (see (\ref{eq:EntStepsRe}) and (\ref{eq:FinalSIneq})),
and next visualized them in Figure \ref{fig:EntropyBounds} as well
as additionally summarized in Subsubsection \ref{subsec:The-entropy-bound}.

In the second part of the work (Section \ref{sec:Some-examples-of}),
for illustration purposes, two multiparametric families of unitary
semicausal qubit toy models have been considered: tensor product models
($3\textrm{Q}^{m}$ models) and controlled non-product models ($l\textrm{C}3m\textrm{Q}$
models). By virtue of the construction of the $3\textrm{Q}$ model
(Figure \ref{fig:T-QModel}), the role played by individual qubits
and consequently the direction of the flow of information are rather
clear. In particular, we can state that ``external information''
carried by the $1^{\textrm{st}}$ qubit (a ``matter'' qubit in the
state $\ket{x}$) is exchanged by the swap gate with that in the outgoing
Hawking qubit (the $3^{\textrm{rd}}$ qubit in the box $U_{0}$).
Therefore, evidently, in the $3\textrm{Q}^{m}$ models, Hawking radiation
produces BH entropy. In turn, at the latter stages of BH evolution,
the accumulated BH entropy is gradually being cancelled by the partner
qubits of the Hawking pairs (see the $2^{\textrm{nd}}$ qubit in box
$U_{2}$). In other words, ``truly external'' (carried by the $1^{\textrm{st}}$
qubit) information does not cross the BH horizon and is (therefore)
not lost. In Subsection \ref{subsec:Non-product-models}, to ``dynamically
close'' the system, we have introduced the controlled non-product
models with a limited number of $l$ (external) control qubits, which
serve to parameterize the Hawking effect. Since the control qubits
are supposed to enter the BH as well as to be later finally released,
in these models, we are forced to (additionally, but hopefully reasonably)
assume that a very small number of particles is unable to support
the existence of the BH, and therefore the BH has to decay and free
the rest of particles (exactly, the control matter particles). Presumably,
such a physics scenario could be conceivable for the total mass of
all $l$ control particles less than the Planck mass, i.e., $M_{l\textrm{\thinspace control}}<M_{\textrm{Planck}}$.
Consequently, in the controlled non-product models, a small contribution
to the BH entropy also comes from the control matter particles.

\bibliographystyle{unsrtnat}
\bibliography{unitary_semicausal}

\begin{acknowledgements}
This work was supported by the University of \L \'{o}d\'{z}.
\end{acknowledgements}

\end{document}